# Temperature induced inversion of oxygen response in CVD graphene on SiO$_2$

Raivo Jaaniso[1,*], Tauno Kahro[1], Jekaterina Kozlova[1], Jaan Aarik[1], Lauri Aarik[1], Harry Alles[1], Aare Floren[1], Alar Gerst[1], Aarne Kasikov[1], Ahti Niilisk[1] and Väino Sammelselg[1,2]

[1]Institute of Physics, University of Tartu, Riia 142, Tartu, 51014, Estonia
[2]Institute of Chemistry, University of Tartu, Ravila 14a, Tartu, 50411, Estonia

**Abstract**

We have synthesized single-layer graphene on Cu foils using chemical vapor deposition method and transferred the graphene to the top of a Si/SiO$_2$ substrate with a pair of prefabricated Ti/Au electrodes. A resistive graphene-based gas sensor prepared in this way revealed n-type oxygen response at room temperature and we have successfully fitted the data obtained with varying oxygen levels using a two-site Langmuir model. P-type oxygen response of our sensor was observed after the temperature was raised to 100 $^o$C, with a reversible transition to n-type behaviour when the temperature was lowered back to room temperature. Such inversion of the gas response type with temperature was interpreted as a result of interplay between the adsorbate-induced charge transfer and charge carrier scattering. The transduction function was derived, which relates the electrical response to surface coverage through both the induced mobility and charge density changes.

---

[*] Corresponding author. Email: raivo.jaaniso@ut.ee, tel.: +3727374731, fax: +372383033

# 1. Introduction

Graphene as a two-dimensional material has every atom at its surface and possesses thus a great potential for the use in chemical sensors. It has been demonstrated, indeed, that using micrometer-size resistive sensors made from exfoliated graphene, single-molecule detection of gases can be achieved [1]. Recent rapid progress in preparation of large-area graphene by thermal treatment of SiC [2, 3] and chemical vapor deposition (CVD) method on transition metals [4, 5] has made this even more realistic [6,7] and serious efforts have been devoted to the studies of various gases like $O_2$, $H_2$, $NH_3$, $NO_2$, CO [8-15]. The lowest detection limits have been achieved for $NO_x$ in the backgrounds of ultrapure inert atmospheres [1, 16]. From the point of view of applications, however, it is important to study the gas response of graphene sensors under ambient conditions. The response to the test gases is then essentially influenced by the interactions of graphene with surrounding atmospheric gases. Overall, these interactions are known to determine the conductivity and the response type (n- or p-type) of the sensor.

Exfoliated and CVD graphene have typically p-type conductivity at ambient conditions because of residual species (e.g. oxygen and water molecules) adsorbed on the surface of graphene in air [11, 17]. At such conditions the conductivity of graphene increases when exposed to oxidising gases, such as $O_2$ [18, 19] or $NO_2$ [9, 20, 21]. The proposed mechanism is electron transfer from graphene to adsorbed molecules, which produces, in case of hole conduction, an increase of free carrier (hole) density. Hence, at ambient conditions the conductivity increases for oxidising gases and decreases for reducing gases, similarly to semiconductor gas sensors. This behaviour ('p-type' response) has been observed in the majority of graphene gas sensor studies at mildly elevated temperatures but also at room temperature. In particular, recently an oxygen sensor has been demonstrated at room temperature with p-type response [19].

Different mechanisms have been proposed for graphene p-doping. Firstly, residual epoxide and carboxylic groups expected in chemically produced graphene are electron-withdrawing and promote holes into the conduction band [8]. Secondly, on $SiO_2$ substrate the p-doping is also believed to originate from adsorbed water molecules, possibly in combination with interactions between these molecules and substrate [11, 22, 23]. The water effect is not uniquely established but one essential mechanism of p-doping in the presence of humidity has been shown to originate from $O_2/H_2O$ redox couple [24]. The roles of water, OH groups and redox pairs have been shown to be certainly important in case of CVD graphene on $SiO_2$/Si substrates.

There are also examples of n-type graphene [25, 26], for which the gas responses are inversed ('n-type'). Graphene on $SiO_2$/Si can be reduced to n-type only after prolonged heating in high vacuum and it quickly returned to the state of hole conductor after exposure to ambient conditions [26].

Epitaxially grown graphene on SiC has shown more stable n-type behaviour due to persistent electron donation from the underlying substrate [17]. It has also shown interesting transition from n-type to p-type behaviour (as a change in gas response direction) with increasing adsorption of electron withdrawing $NO_2$ gas [25].

In this paper we report on experiments with resistive gas sensors based on CVD graphene, which showed n-type oxygen response at room temperature and p-type oxygen response at elevated temperatures. The samples were used without any special pre-treatments in high vacuum or at high temperature, which allowed conducting the experiments closer to real conditions for gas sensors. The results, measured at different oxygen concentrations, were successfully modelled and the response inversion with temperature was interpreted as a result of interplay between the adsorbate-induced charge transfer and charge carrier scattering.

## 2. Experimental

### 2.1. Graphene growth and sensor fabrication

The monolayer graphene was grown on commercial 25-μm thick polycrystalline copper foils (99.999%, Alfa Aesar) in a laboratory-made CVD reactor following the receipt described in [27]. First, the foils were annealed for 30 minutes in $Ar/H_2$ (both 99,999%, AS AGA Estonia) flow of 100/120 sccm (standard cubic centimeters per minute) at a temperature of 950 $^o$C and after that they were exposed to a gas mixture of 10% $CH_4$ (99,999%, AS AGA Estonia) in Ar at the same temperature for about 10 minutes. The flow of the $CH_4$/Ar gas was kept at 40 sccm. The graphene samples on copper foils were then cooled to room temperature with a rate of about 15 $^o$C/min in an Ar-gas flow of 100 sccm. Next, the upper sides of copper foils with graphene were covered with polymethyl methacrylate (PMMA; M~997 000, GPC, Alfa Aesar) in chlorobenzene (Alfa Aesar) solution (20 mg/ml) and copper was removed in warm $FeCl_3$ (97%, Alfa Aesar) solution (1 mol/l). The PMMA/graphene structures were then washed with deionized water and transferred to the top of $Si/SiO_2$ substrates with Ti/Au electrodes (3 nm/60 nm) that had been deposited through the shadow mask by electron beam evaporation (see Fig. 1a). PMMA was dissolved by dichloromethane (Alfa Aesar). Finally, the samples were washed in hot acetone (99.5%, Carl Roth GmbH&Co). The optical microscope image of one of gas sensors prepared in this way is shown in Fig. 1b. Characterisation by SEM showed that the electrode gap was continuously covered by graphene film. In Fig. 2a inset, a SEM image of the same sensor is shown: within otherwise continuous area a grain boundary and some defects can be seen.

## 2.2. Characterization of gas sensors and gas sensitivity measurements

After fabrication our graphene-based gas sensors were held in ambient air for several weeks and we did not perform any heat treatment with them before the gas sensitivity measurements. The quality of graphene was studied by a Raman spectrometer (Renishaw inVia) with laser excitation of 514.5 nm. In addition, electron probe microanalysis (EPMA) method was applied to characterize the graphene films between the electrodes of the sensors. EPMA measurements were performed using INCA Energy 350 (Oxford Instruments) energy dispersive x-ray spectrometer (EDS) connected to scanning dual beam microscope Helios$^{TM}$ NanoLab 660 (Fei) after the gas sensitivity measurements. The measurements of the gas sensitivity were carried out with a sourcemeter (Keithley 2400), gas mixing system based on three mass flow controllers (Brooks, model SLA5820), and a sample chamber with a small hotplate heater. The voltage applied between the two Au-electrodes (with a gap of about 100 μm) was typically 100 mV. The gases used in our measurements, $N_2$ and $O_2$, were both 99.999% pure (AS AGA Estonia). The gas flow through the sample chamber was kept constant at 200 sccm but the ratio between the flow rates of two gases was varied for changing the oxygen content. Before the measurements the samples were always held under constant voltage in the flow of synthetic air (21% of oxygen, 79% nitrogen) for about an hour.

## 3. Results

### 3.1. Raman spectra

Figure 2a shows a typical Raman spectrum from graphene between the electrodes of a gas sensor depicted in Fig. 1b. The locations of G- and 2D-peaks at ~1585 cm$^{-1}$ and ~2687 cm$^{-1}$ and their full widths at half maximums of 15 cm$^{-1}$ and 35 cm$^{-1}$, correspondingly, are characteristic to a single layer graphene [28]. Over 100 spectra, taken from random locations within the electrode gap, were normally recorded for characterising the samples. Over 90% of these spectra showed characteristic features of monolayer graphene: (i) a symmetric 2D-band centred around 2687 cm$^{-1}$ with full width at half maximum of 30-40 cm$^{-1}$; (ii) approximately 1:3 ratio of G- and 2D-band peak intensities. The absence or low intensity of D-band (~1350 cm$^{-1}$) indicated low defect density. The rest (~5%) of the Raman spectra had different characteristics and could not be assigned to the single layer graphene. A correlation between the positions of G-peak [Pos(G)] and 2D-peak [Pos(2D)] was also studied for our samples. The positions of the peaks are known to vary slightly because of local disorder and interactions with the substrate. However, if the positions are correlated, then the sign of correlation can be used as an indicator of the conductivity type [29]. The typical data are plotted in Fig. 2b. The linear regression of points yields a positive slope (equal to $0.25 \pm 0.08$; respective fitting result is shown in Fig. 2b by dashed line), whereas the correlation coefficient between Pos(G) and Pos(2D) is equal to 0.24. If one disregards five somewhat anomalous points with the smallest G-band frequencies in Fig. 2b, then the correlation coefficient is even 0.5. The positive correlation refers to the p-doped graphene [29] as is typical for graphene devices at ambient conditions.

### 3.2. Gas response of sensors

In Figure 3, the responses of the gas sensor depicted in Fig. 1b are shown when the gas composition has been switched between synthetic air and pure nitrogen. As one can see, the relative change of the signal of our sensor is of the order of 1% and it is fully reproducible with a baseline slightly drifting upwards. Note that the conductivity decreases when oxygen is introduced into the gas, which refers to the n-type conductivity of our sensor because the adsorbed oxygen is generally acting as an electron acceptor. This finding is unexpected because graphene is generally a p-type material under ambient conditions due to doping from water and oxygen [11,17] and also our data on Raman mapping of the graphene between electrodes (Fig. 2b) refer to the p-type material.

In order to have a closer look at such anomalous oxygen sensitivity, the responses at different oxygen levels were studied. In the series of measurements, shown in Fig. 4, the oxygen content was switched to different levels and always retuned to air (reference) level before the next change. The response is defined here and afterwards as a relative change of conductance $G$:

$$S = \frac{G - G_0}{G_0}, \tag{1}$$

where $G_0$ is the conductance in air.

From the data in Fig. 4, the stationary values of conductance at each oxygen level were obtained (except for the 0%-level, when the conductance did not reach the stationary state) and respective stationary responses were plotted against the oxygen content (see Fig. 5).

This overall stationary response in Fig. 5 (shown with crosses) could be successfully fitted with a ratiometric function, characteristic for a Langmuir isotherm ([30], see Appendix):

$$S_S = c \frac{x - x_0}{1 + bx}, \tag{2}$$

where $b$ is the affinity constant and $x_0 = 21\%$. In case of a Langmuir-type processes one may expect the temporal response after a step-wise change in oxygen content to be exponential, with a rate constant given by the sum of desorption ($K_D$) and adsorption ($k_A x$) rates ($x$ is the oxygen level introduced at a step-change, $k_A$ is constant):

$$k = K_D + k_A x. \qquad (3)$$

This hypothesis was tested with all transient curves depicted in Fig. 4. It turned out that the temporal responses could not be fitted by a single exponential but in all cases followed exceptionally well a double exponential function:

$$S(t) = A_0 + A_1 \exp[-k_1(t-t_0)] + A_2 \exp[-k_2(t-t_0)], \qquad (4)$$

where $t_0$ is the initial moment of time, when the respective step-wise change in gas composition was introduced. An example of fitting is given in semi-logarithmic scale in Fig. 6, where the experimental data are representing the change from 10% to 21% oxygen content (it is the same curve, which starts at $t_0 = 2005$ s in Fig. 4).

The rate constants, $k_1$ and $k_2$, determined from fitting of all transient curves present in Fig. 4, are shown in Fig. 7 as the functions of oxygen content. As one can see, both rate constants follow reasonably well the linear dependence of Eq. (3).

The amplitudes of two components, $A_1$ and $A_2$, as found from fitting the transient curves, are shown together with approximations by Eq. (2) in Fig. 5. All parameters of two components (sites), obtained by fitting the results in Figs. 5 and 7, are collected into Table I.

Table I. Parameters of two adsorption sites, describing the oxygen response at room temperature.

| Parameter | Site I | Site II |
|---|---|---|
| $k_A$ (1/%s) | $0.0034 \pm 0.0008$ | $0.00099 \pm 0.00008$ |
| $K_D$ (1/s) | $0.093 \pm 0.015$ | $0.0028 \pm 0.0015$ |
| $b$ (%) | $0.037 \pm 0.010$ | $0.35 \pm 0.18$ |
| $c$ (%) | $-0.033 \pm 0.003$ | $-0.041 \pm 0.006$ |
| $\alpha$ (%) | $-1.6 \pm 0.2$ | $-0.98 \pm 0.15$ |

After room temperature experiments we also studied the response of our sensor at elevated temperatures. After raising the temperature to 100 °C and allowing the stabilisation for about an hour in synthetic air, a similar switching of the gas composition between air and nitrogen was performed as at room temperature. As one can see from Fig. 8, a significantly larger response (~15%) with an opposite sign was observed at 100 °C. Such response is in accordance with the influence of oxidizing gas on p-type material: the adsorbed species as hole donors increase the density of charge carriers and hence the conductivity of the sensor. After the temperature was stepwise lowered back to room temperature, the response of the sensor was also recorded at intermediate temperatures. It can be seen from Fig. 8 that the response was again inversed and the switchover between two response types occurs between 50 and 75 °C. After lowering the temperature back to room temperature (25 °C), the initial n-type response of the sensor was recovered, although with somewhat smaller amplitude.

### 3.3. Modelling of the response

Our results on oxygen sensitivity at room temperature can be explained with a following model. If one assumes the presence of a single (major) type of charge carriers, the conductivity can be written as

$$\sigma = en\mu, \tag{5}$$

where $e$, $n$ and $\mu$ are the charge, density and mobility of charge carriers. When the gas adsorption induces the carrier density change $\Delta n$ and their mobility change $\Delta \mu$, then the conductivity change is

$$\Delta\sigma = e(n+\Delta n)(\mu+\Delta\mu) - en\mu = \sigma\left(\frac{\Delta\mu}{\mu} + \frac{\Delta n}{n} + \frac{\Delta n}{n}\frac{\Delta\mu}{\mu}\right). \tag{6}$$

Further we assume that $\Delta n$ is proportional to the fractional coverage change $\Delta\theta$ by the adsorbed species:

$$\Delta n = \tilde{n}\Delta\theta, \tag{7}$$

where the constant $\tilde{n}$ characterises the change of carrier density per unit change of coverage. Equation (7) can obviously be used up to the densities of adsorbed species when their interactions and screening effects start to appear. In the approximation of the Matthiessen's rule the mobility can be written as (when also neglecting a possible weak dependence from charge density) [31, 32]:

$$\frac{1}{\mu} = \frac{1}{\mu_0} + \frac{\theta}{\mu_1}, \tag{8}$$

where the mobility $\mu_0$ is determined by interactions with phonons, other impurities or disordered features in the absence of adsorbed gas molecules; $\mu_1$ is a constant, the inverse value of which describes the scattering strength by the latter. The relative change of mobility at a coverage change $\Delta\theta$ follows from Eq. (8):

$$\frac{\Delta\mu}{\mu} = -\frac{\mu_0 \Delta\theta}{\mu_1 + \mu_0(\theta + \Delta\theta)}, \tag{9}$$

where $\theta$ is the initial coverage. After inserting Eqs. (7) and (9) into Eq. (6) one obtains for the relative conductance (and conductivity) change:

$$S = \frac{\Delta G}{G} = \frac{\Delta\sigma}{\sigma} = \left[\frac{\tilde{n}}{n}\left(1 + \theta\frac{\mu_0}{\mu_1}\right) - \frac{\mu_0}{\mu_1}\right]\frac{\Delta\theta}{1 + \frac{\mu_0}{\mu_1}(\theta + \Delta\theta)}. \tag{10}$$

In a special case $\mu_1 \gg \mu_0$ the mobility change introduced by adsorbing species is relatively small and the general transducer function, Eq. (10), simplifies to

$$S = \left(\frac{\tilde{n}}{n} - \frac{\mu_0}{\mu_1}\right)\Delta\theta = \alpha\Delta\theta. \tag{11}$$

The sign of transduction coefficient $\alpha$ in Eq. (11) can be positive or negative, it depends not only on the type of conductivity (n- or p-type) and gas involved (oxidising or reducing) but also on the balance of the relative changes of carrier density and mobility, as well as on the sign of the mobility effect ($\mu_1$ can be positive or negative depending on whether the adsorbing species decrease or increase the mobility [33, 34]).

In case of p-type material and oxidizing gas $\tilde{n} > 0$, and therefore the only possibility to have a negative transduction coefficient $\alpha$ as observed in our room temperature experiments is when the mobility is decreasing with increasing coverage [32, 35] and the mobility change dominates over the carrier density change, i.e., when $\mu_1 > 0$ and

$$\frac{\tilde{n}}{n} < \frac{\mu_0}{\mu_1} \quad . \tag{12}$$

The part of the model describing the receptor function, i.e., the dependence of the coverage factor from the gas pressure (or content) is based on Langmuir adsorption-desorption kinetics and is given in the Appendix.

The final result is that if one assumes the presence of two adsorption sites for oxygen, the model (uniting both the receptor and transducer functions) leads explicitly to Eqs. (2)-(4), which were used at fitting the experimental data. In other words, our experimental results can be successfully (at a precise quantitative level) explained by two-site Langmuir sorption, transduced into electrical response by function (11). The anomalous (negative) transduction coefficient $\alpha$ at room temperature can be explained by the dominance of mobility change at oxygen adsorption. Note that $\alpha$ is related to parameters $c$ and $b$ by Eq. (A5) and that the values of $\alpha$ are given in the last row of Table I. The inversed response at high temperatures, over 75 $^\circ$C, can be explained by the increased role of charge transfer at oxygen adsorption, which inverses the relation (12) and results in positive transduction coefficient.

## 4. Discussion

Gas sensitivity of resistive graphene sensors is usually explained [6, 7] as a result of charge transfer taking place at adsorption-desorption processes, which leads to the change of free carrier density and hence conductivity of graphene. In the pioneering paper [1], where the gas sensitivity mechanisms were studied, no any effect of gas sorption on the mobility of carriers was found.

However, it is difficult if not impossible to explain our results with charge transfer mechanism solely. In this case we have to assume that the sample has initially, at room temperature, n-type conductivity. Then, by introducing oxygen into the measurement chamber, its adsorption on graphene leads to the depletion of electrons from the sample and hence to the decrease of conductance, as seen in Fig. 3. N-type conductivity of graphene has been observed on different substrates but it has been stabilized at ambient conditions only on SiC [17, 25]. While it has been demonstrated in Ref. [26] that initially p-doped (exfoliated) graphene on Si/SiO$_2$ substrate can be changed to n-type after ~20 h of annealing in vacuum, the p-doping was quickly restored in ambient conditions. Moreover, we did not perform any special treatments (heating over 100 $^{o}$C, keeping in high vacuum) before the gas sensitivity measurements.

Further, even if we have an exotic sample with, e.g., an impurity-induced electron doping, one still has to explain the transfer of conductivity type (form n-type to p-type) with increasing temperature. In principle, such a transfer can occur as a result of doping by the test gas. Some examples can be found in case of $NO_2$ gas [25, 36], when initially n-type samples were converted into p-type after exposure to strongly oxidising gas. In these examples, the adsorption of an oxidant ($NO_2$) depletes the samples strongly from the electrons leading to the change majority carriers. In our case, one rather induces the reduction in the number of hole dopants, as with increasing temperature the coverage of adsorbed oxygen (and OH groups, if present) should decrease. Therefore, it seems possible to explain the transfer from p-type conductivity to n-type with increasing temperature in case of graphene on $Si/SiO_2$ but not vice versa.

Regarding the role of mobility, several recent studies have demonstrated that the adsorption of ambient gases like oxygen and water vapor can lead to scattering of charge carriers and change the mobility at least up to 50% [32, 34]. Also, the cleaning of graphene after storage in ambient conditions by heating in vacuum has been shown to enhance the mobility by several times [35, 37]. Consequently, there is clear evidence that generally the transducer function given by Eq. (10) or Eq. (11) has to be used for describing the gas responses of graphene. Note that in our samples the observed response was relatively small at room temperature, in the order of 1%. Such change can be produced by a minute change of mobility and hence the approximation $\mu_1 >> \mu_0$, used in the derivation of Eq. (11) is well justified.

Regarding the interpretation of response inversion at the end of Section 3, the main remaining question is: why the mobility change dominates over the carrier density change at room temperature, and why this condition is inversed after heating the sample to 100 $^{o}$C? Taking into the account the arguments given above, the following explanation can be offered: the adsorption sites, responsible for mobility change and charge density change are different, the latter ones being passivated at room temperature. We note here, that not all the studied samples were sensitive to oxygen at room temperature. For a number of samples the response was significantly smaller (0.1% or less, i.e. practically nonexistent) at room temperature but had approximately the same value (10-15%) and (p-)type at 100 $^{o}$C. Obviously such samples were completely passivated at room temperature and did not contain any oxygen adsorption sites influencing the carrier scattering properties.

The fact that the processes taking place at room temperature and at 100 $^{o}$C have different nature is also confirmed by comparing the rates of transient responses at these temperatures, respectively. The responses at 100 $^{o}$C were not studied in detail but the curves, depicted in Fig. 9, could also be fitted well with Eq. (4). The overall process was, however, quite slower at 100 °C. For example, the fitting of the transient response at changing the gas composition from air to nitrogen resulted for the faster component $k_1 = 0.016$ s$^{-1}$ as compared to the respective value $k_1 = 0.13$ s$^{-1}$ at 23 °C. Obviously these two $k_1$ values characterise different mechanisms, otherwise one should suppose decreasing desorption rate with increasing temperature.

When comparing the parameters of two sites in Table I, following observations can be made:
1) one site has 30 times bigger desorption rate and 3 times bigger absorption rate;
2) transduction coefficient is negative for both sites and not so different (1.6 times) in the absolute value.

Consequently, both sites influence the conductivity in a comparable manner but the chemical interaction is rather different. At this stage, one can only speculate about the origin of the sites, which can be related either to the edges of graphene [18, 38], impurities, or SiO$_2$ substrate [11]. The impurities may only be present due to unintentional chemical doping during CVD or graphene transfer process. We believe that the most likely sources for such doping are the residuals of FeCl$_3$ solution which was used to etch away the copper foils and these residuals might have remained sealed between SiO$_2$ substrate and graphene after the graphene/PMMA structures were lifted to the top of the prefabricated sensor electrodes. This conclusion bases on EPMA analysis, which revealed the presence of small amount (< 1%) of impurities from FeCl$_3$ etching solution in the graphene between the electrodes of the sensor depicted in Fig. 1b.

The passivation of the sites, which become active at approaching to 100 °C, may be related to the residual water or OH groups. The presence of such species cannot be ruled out, because although we used nominally dry gases in the experiments, the samples (as well as the gas tubes and chamber) were not specially heated or dried. Graphene, of course, is known to be highly hydrophopic but it has been demonstrated that water can easily adsorb onto the edges of graphene sheets [18, 38] and on SiO$_2$ substrate [23, 39]. The presence of residual humidity can also explain the anomalous temperature dependence of conductivity in nitrogen environment: as one can see in Fig. 8, the steady-state conductance in N$_2$ is smaller at 100 °C than at 50 and 75 °C. It has been shown previously that humidity acts as electron acceptor on graphene [21]. It follows from this fact that effective removal of water molecules (or related species) at 100 °C should lead to the decrease of conductivity in p-type material. In other words, the upper curve in Fig. 8, taken at 100 °C, is lowered with respect to other curves due to removal of residual humidity from the graphene surface. From the almost perfect two-exponential transient responses (Fig. 6) and the model outlined in Section 3 and Appendix one may conclude that two well-defined adsorption centres exist for oxygen. An essential argument for the validity of the model is that the affinity parameters (*b*) were independently determined from the kinetic constants and from the amplitudes of two exponentials, whereas the results coincided within the error margins given in Table I. Even if the exact nature of the sites remains unclear, we have demonstrated that graphene can be a perfect model system with limited heterogeneity. We completely agree with the conclusion of Ref. [32] that graphene offers simple model platform for the studies of host-guest interactions and sorption mechanisms.

## 5. Conclusion

To conclude, we have prepared and studied the oxygen response of resistive gas sensors made from CVD graphene on $SiO_2$/Si substrates. At room temperature, an n-type oxygen response (increasing conductivity with increasing oxygen concentration) was observed. For a p-type, hole conducting material such gas response cannot be explained as arising from electron capture at oxygen adsorption. However, it can be interpreted by assuming that the adsorbed oxygen induces additional charge carrier scattering. This leads to the conductivity decrease with increasing oxygen content, in agreement with the experiment. The conductivity type of the samples was confirmed by extensive Raman-mapping experiments. When raising the temperature up to 100 $^o$C, a p-type response was observed for oxygen, which could be explained by conventional charge transfer mechanism. For uniting the interpretations at two different temperatures a general transduction function was derived, which relates the electrical response to fractional surface coverage and involves the effect of adsorbed species on both the charge carrier density and mobility. The dependence of gas response on oxygen concentration was measured at room temperature and modelled with the help of transducer function and two-site Langmuir receptor function. Good quantitative agreement between the experimental data and the model allowed detailed determination of desorption and adsorption parameters.

*Acknowledgements*

This study was supported by the European Union through the European Social fund (Grant MTT1) and the European Regional Development Fund (project "Mesosystems: theory and applications", 3.2.0101.11-0029). We also acknowledge financial support from Estonian Research Council (grant IUT2-24 "Thin-film structures for nanoelectronic applications and functional coatings") and from Estonian Center of Excellence in Research "High-technology Materials for Sustainable Development" (TK117T).

## APPENDIX: Model for the receptor function

If the coverage factor $\theta(t)$ is governed by the first order Langmuir kinetics, then

$$\Delta\theta(t) = \theta(t) - \theta_i = (\theta_f - \theta_i)\{1 - \exp[-(Pk_A' + K_D)t]\}, \tag{A1}$$

where $\theta_i$ is the initial coverage (at $t = 0$), $P$ is the gas pressure introduced at $t = 0$, and $\theta_f$ is the final stationary coverage

$$\theta_f = \frac{Pk_A'}{Pk_A' + K_D}. \tag{A2}$$

For comparison with the experimental results we introduce the gas content in air, $x$ (in %), instead of partial pressure $P$:

$$x = 100\frac{P}{P_{air}}. \tag{A3}$$

Then $Pk_A'$ has to be replaced by $xk_A$ in Eqs. (A1) and (A2). With this replacement, the rate constant in the exponent of Eq. (A1) is given by Eq. (3) and the stationary gas response relative to a certain reference composition $x_0$ can be derived from Eq. (11) and Eq. (A2) as follows:

$$S_S = \frac{G - G_0}{G_0} = \alpha\left(\frac{xk_A}{xk_A + K_D} - \frac{x_0 k_A}{x_0 k_A + K_D}\right) = \frac{\alpha k_A K_D}{x_0 k_A + K_D}\frac{x - x_0}{xk_A + K_D}. \tag{A4}$$

It is easy to see, that the latter result coincides with Eq. (2) when

$$c = \alpha\frac{b}{1 + bx_0}, \tag{A5}$$

and

$$b = \frac{k_A}{K_D}. \tag{A6}$$

Finally, in the case of two independent sorption sites (I and II) one has instead of Eq. (11) under a similar assumption of relatively small mobility change:

$$S = \frac{\Delta\sigma}{\sigma} = \alpha_I \Delta\theta_I + \alpha_{II} \Delta\theta_{II} \ . \tag{A7}$$

Both coverage changes in Eq. (A7) are given by Eq. (A1) but with different parameter values characteristic for two sites. Consequently, Eq. (A7) is given by a sum of two exponentials as Eq. (4), whereas the amplitudes in the latter equation are related to the basic model parameters through a double set of Eqs. (A2)-(A6).

**Biographies**

**Raivo Jaaniso** received his PhD in solid-state physics from the Institute of Physics, Estonian Academy of Sciences in 1988, and holds currently a position of senior scientist at the University of Tartu. His research interests have covered a wide area from site-selective laser spectroscopy and spectral hole burning on both organic and inorganic (rare-earth doped) materials to pulsed laser deposition of functional thin films and investigations of luminescent and semiconductor gas sensor materials.

**Tauno Kahro** received his MSc degree in materials science from the University of Tartu in 2011. He is currently studying for a Ph.D. degree in the same university. His research is focused on electrical characterization of graphene-based nanostructures.

**Jekaterina Kozlova** received her MSc degree in inorganic chemistry from the University of Tartu in 2009. She is currently studying for a Ph.D. degree in the same university. Her research interests are carbon materials and thin film characterization.

**Jaan Aarik** received his BSc degree from the University of Tartu in 1974, MSc degree from the University of Tartu in 1994 and PhD from the University of Tartu in 2007. Since 2009 he is a professor at the Institute of Physics of the University of Tartu. His main research activities have been related to investigation of diode lasers and photodiodes, development of thin-film deposition methods and equipment, and characterization surface reactions on thin solid films.

**Lauri Aarik** started PhD studies in materials science at the beginning of 2012 and works as an engineer at the University of Tartu. His is specialized in making the ultrathin organic and inorganic anticorrosive layers and developing atomic layer deposition and chemical vapour deposition reactors for different applications.

**Harry Alles** received his MSc degree from the University of Tartu in 1990 and PhD degree from Helsinki University of Technology in 1995. After working as a research associate in the University of Manchester in 1995-1998 he moved back to Helsinki University of Technology being there in a position of a senior scientist for eleven years. In 2009 he returned to the University of Tartu where he is a senior scientist. His current research interests lie mainly in synthesis and characterization of graphene-based nanostructures.

**Aare Floren** received his MSc degree from the University of Tartu in 2005. Currently he is working as an engineer in the Institute of Physics of the University of Tartu and is finalizing his PhD studies. His research interests are centred around oxygen sensors materials, based on luminescence quenching as well as on conductometric principle.

**Alar Gerst** graduated from the University of Tartu in 1961 and since this time has been engaged in the Institute of Physics, University of Tartu (former Institute of Physics, Estonian Academy of Sciences) as a research scientist in investigations of electric and optical properties of different semiconductor structures including heterolasers and photo-resistors. During the last decade he has been involved as an engineer in the investigation of electric and gas sensing properties of thin oxide films and graphene-based nanostructures.

**Aarne Kasikov** received his MSc and PhD in applied physics from the University of Tartu, 1995 and 2010, correspondingly, and is currently a researcher in the Institute of Physics at the University of Tartu. His scientific interests lie in vacuum evaporation and optical measurements of thin films.

**Ahti Niilisk** received his PhD from the University of Tartu in 1971, and has currently a position of senior scientist in the Institute of Physics at the University of Tartu. His scientific interests are ranging from high-pressure spectroscopy of solids to the technology and Raman characterization of thin oxide films and graphene.

**Väino Sammelselg** received his PhD degree in solid state physics (optics) from the Institute of Physics, Estonian Academy of Sciences in 1989 and holds currently positions of professor of inorganic chemistry in the Institute of Chemistry and head of department of materials science in the Institute of Physics, both in Faculty of Science and Technology of the University of Tartu. He has been engaged in different fields of materials studies, beginning with mixed ion crystals, $A^3B^5$ semiconductor heterostructures and glass-fiber structures. His current research is focused on the metal oxides, conductive polymers and graphene thin films – on their preparation and nano-characterization, and on the solid surfaces functionalization, including passivation and protection against corrosion.

**Figures and Figure Captions**

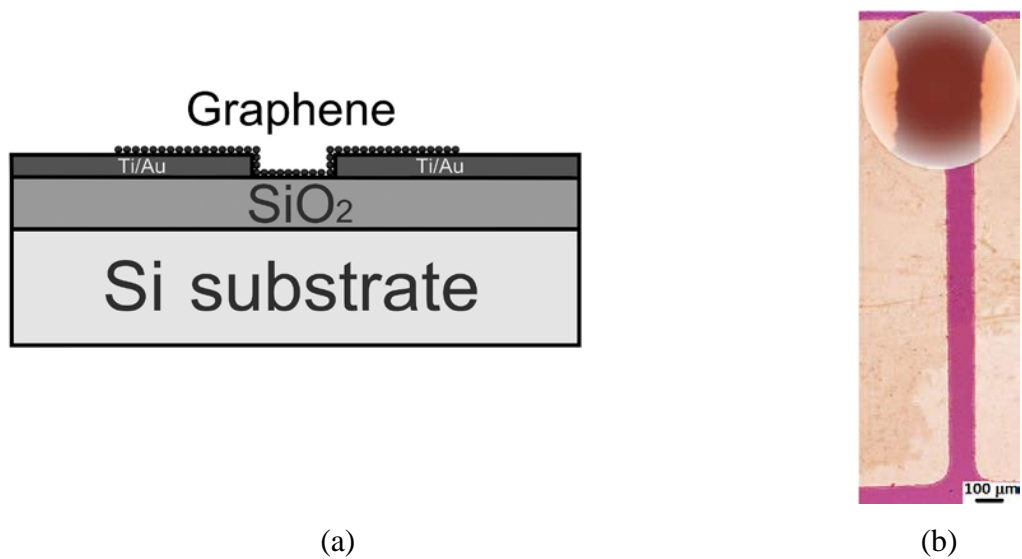

(a)            (b)

**Fig. 1.** (color online) (a) Schematic illustration of our gas sensors based on CVD graphene. (b) Optical microscope image of one of our sensor structures. A few times magnified view of the underlying part of the sensor is given on the top.

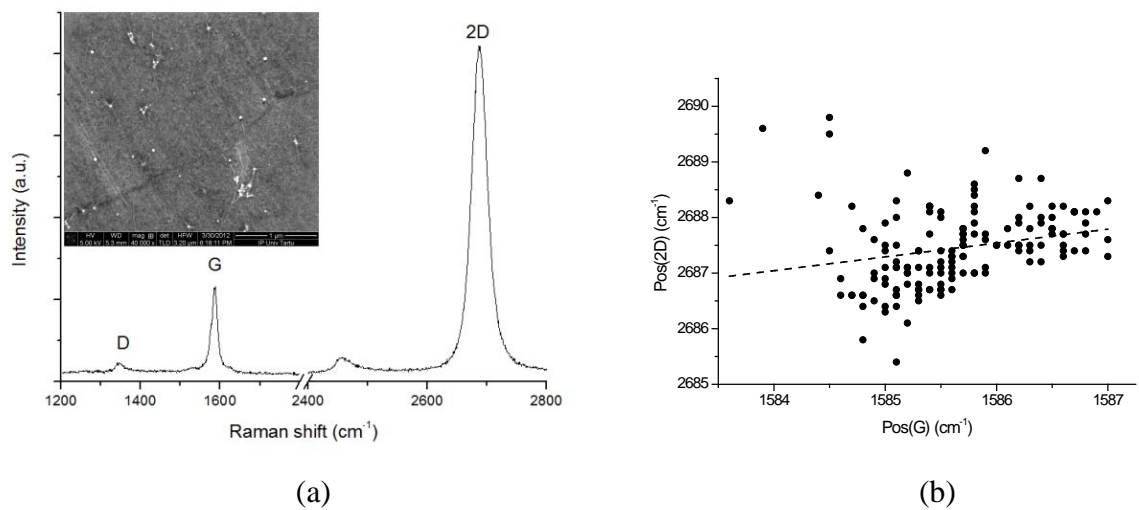

(a)            (b)

**Fig. 2.** (a) Typical Raman spectrum and SEM image (inset) of CVD graphene between the electrodes of the sensor depicted in Fig. 1b. The size of SEM image is $2.3 \times 3.2$ μm. (b) Data of Raman mapping [Pos(G) and Pos(2D) indicate the positions of G- and 2D-peaks, respectively, which were extracted from altogether 150 spectra of graphene between the electrodes of the sensor in Fig. 1b].

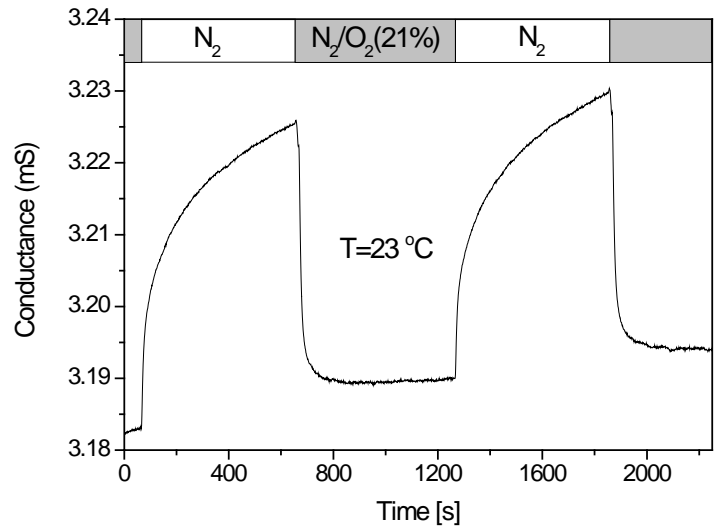

**Fig. 3.** Sensor responses to changes of gas composition (nitrogen and synthetic air) at room temperature.

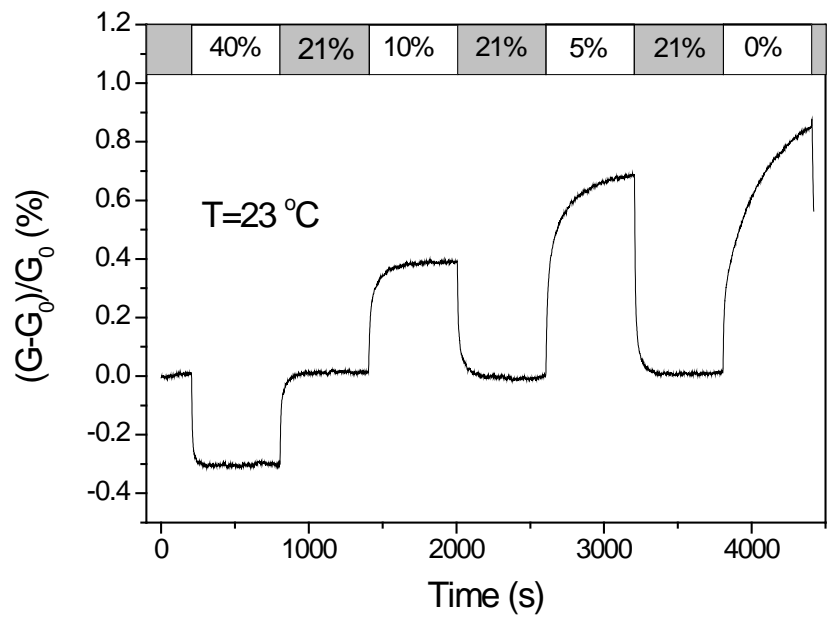

**Fig. 4.** Gas sensor relative responses to the varying amount of oxygen in the $N_2/O_2$ mixture. The oxygen content in the mixture is shown at the upper bar.

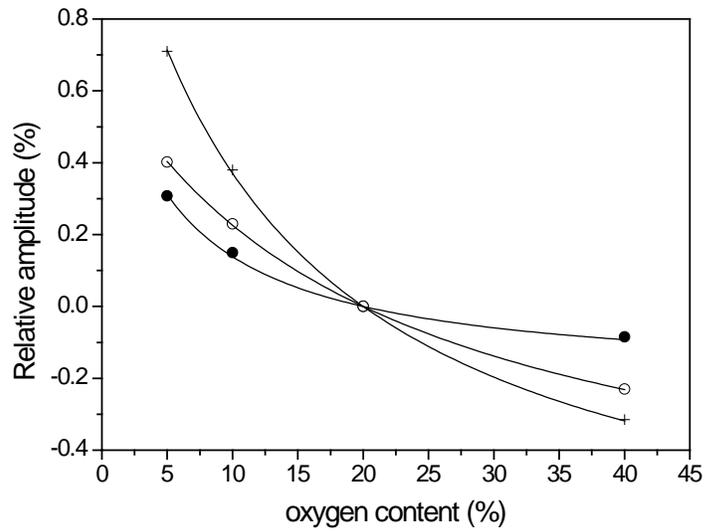

**Fig. 5.** Stationary gas response as a function of oxygen content in $N_2/O_2$ mixture (crosses). The amplitudes of two components $A_1$ (empty circles) and $A_2$ (filled circles) are also shown. The fitting curves with Eq. (2) are given with solid lines.

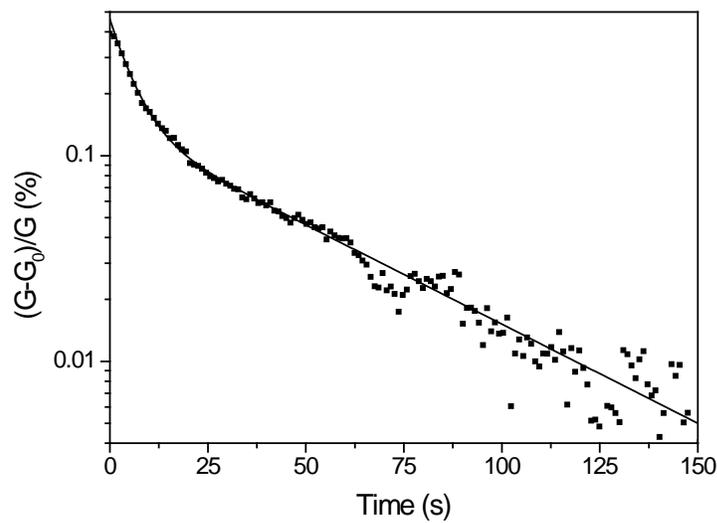

**Fig. 6.** Transient gas response after step-change of oxygen content from 10% to 21% (square dots) and the double-exponential fitting curve (solid line).

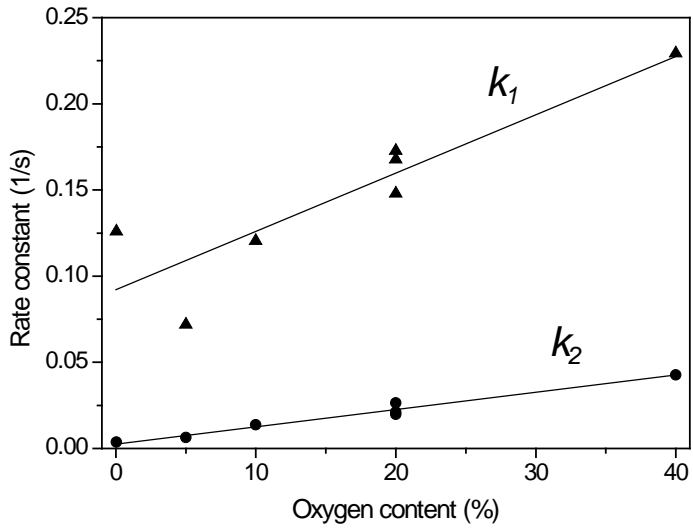

**Fig. 7.** Dependences of two rate constants on oxygen content and their linear fits.

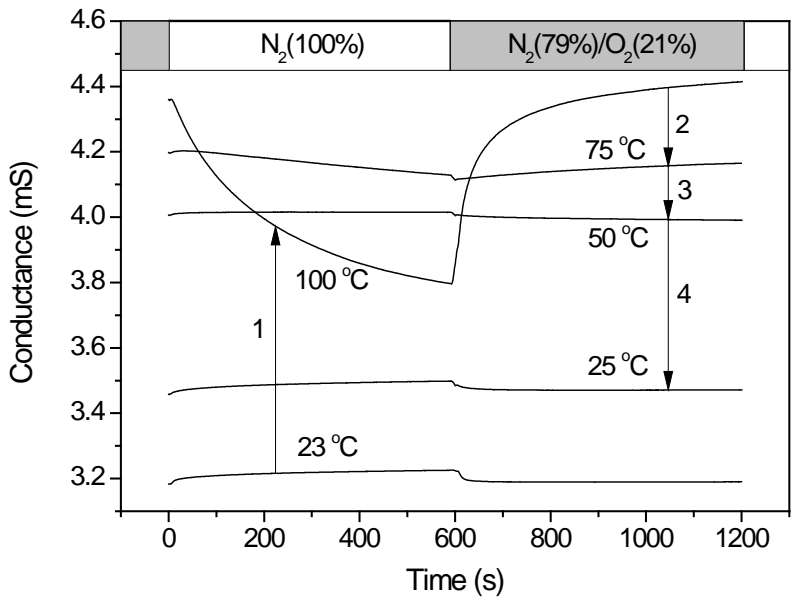

**Fig. 8.** Temperature dependence of gas sensor response to changes of gas composition (nitrogen and synthetic air). The arrows (marked with 1 to 4) show the order in time how the temperature was changed.